\def\({\left(}
\def\[{\left[}
\def\l\{{\left\{}
\def\){\right)}
\def\]{\right]}
\def\r\}{\right\}}
\def\what{\widehat}
\def\raw{\rightarrow}
\def\bA{\bar A}
\def\am{\bar A_{\rm max}}
\def\cf{{\cal F}}
\newcommand{\pd}[2]{\frac{\partial #1}{\partial #2}}

\def\etal{{\sl et al.} }
\def\ratio#1#2{{{#1}\over{#2}}}
\def\KH{Kelvin-Helmholtz }
\def\3d{three-dimensional }
\def\2d{two-dimensional }
\def\be{\begin{equation}}
\def\ee{\end{equation}}
\def\bef{\begin{figure}}
\def\ef{\end{figure}}

\def\DXDYCZ#1#2#3{\left({\partial#1\over\partial#2}\right)_{#3}}
\def\lapp{\mathbin{\raise2pt \hbox{$<$} \hskip-9pt \lower4pt
\hbox{$\sim$}}}
\def\gapp{\mathbin{\raise2pt \hbox{$>$} \hskip-9pt \lower4pt
\hbox{$\sim$}}}

\documentclass{aa}
\usepackage{graphicx}
\begin{document}

   \title{Time-dependent MHD shocks and line emission: The case of the DG Tau jet
}

  \author{
  S. Massaglia\inst{1}
\and A. Mignone\inst{2} \and G. Bodo\inst{2}
}
\offprints{S. Massaglia}
\institute{
Dipartimento di Fisica Generale dell'Universit\`a,
Via Pietro Giuria 1, I-10125 Torino, Italy\\
 email: massaglia@ph.unito.it \and
Osservatorio Astronomico di Torino, Viale Osservatorio 20, I-10025
Pino Torinese, Italy\\
 email: mignone@to.astro.it, 
 email: bodo@to.astro.it
 }
\titlerunning{}
   \date{Received; accepted }

\abstract{ The line emission from a growing number of Herbig-Haro jets can be observed and resolved at angular distances
smaller than a few arcseconds from the central source. The interpretation of this emission is problematic, since the simplest model
of a cooling jet cannot sustain it. It has been suggested that what one actually observes are shocked regions with a
filling factor of $\sim 1\%$. In this framework, up to now, comparisons with observations have been based on stationary shock models. Here
we introduce for the first time the self-consistent  dynamics of such shocks and we show that considering their properties 
at different times, i.e. locations, we can reproduce observational data of the DG Tau microjet.
In particular, we can interpret the spatial behavior of the [SII]6716/6731 and [NII]/[OI]6583/6300 
line intensity ratios adopting a set of physical parameters that yield values of mass loss rates
and magnetic fields consistent with previous estimates. We also obtain the values of the mean
ionization fraction and electron density along the jet and compare these values with those
derived from observations using the sulfur doublet to constrain the electron density.
}
\maketitle
\section{Introduction}
Herbig-Haro jets can be observed at very high spatial resolution by HST and
with the next generation of ground based optical telescopes such as VLTI/AMBER the
angular resolution capabilities will be boosted from
the actual fraction of an arcsecond up to the milliarcsecond range (Bacciotti 2004). It will soon be possible to look into the very first part of the jet as it emerges from the accretion
disk or from the reflection nebula and resolve many jets in their radial extent. It will 
thus be feasible to perform a comparison of direct observations with the predictions of the acceleration
models. Some jets (sometimes called `microjets'), e.g. HH 30 (Bacciotti et al. 1999) or DG Tau 
(Lavalley et al. 1997, Lavalley-Fouquet et al.
2000 (L-FCD2000), Bacciotti et al. 2000), are particularly good candidates for a
high resolution study of the evolution of the physical parameters along 
the initial fraction of the jet, i.e. up to $\sim 10^{16}$ cm where the conditions to meet for
the line emission mechanisms at work are the most severe. 

The jet of DG Tau has been observed close to
its base at CFHT at high angular resolution  ($\sim 0^{\prime \prime}$.5) by L-FCD2000
who showed the behavior of [SII]6716/6731 and [NII]/[OI]6583/6300 
line ratios along the
jet for the high (HV), intermediate (IV) and low (LV) velocity components. 
 Our calculations will address these data in modeling the jet line emission.

These observations typically show that the behavior of temperature, ionization and density 
along the jet is incompatible with a freely cooling jet.
Various heating processes have been proposed in the literature, such
as ambipolar diffusion (Garcia et al. 2001a,b), photoionization by soft X-rays from the
TTauri star (e.g. Shang et al. 2002) and mechanical heating (O'Brien et al. 2003, Shang 
et al. 2002, for X-wind jets). These estimates where carried out
for steady-state jet models and pointed out that mechanical heating was the
most effective in reproducing the observations. The idea of tapping a small fraction
of the jets kinetic energy to convert into heat is certainly appealing; however
 there is no physical explanation for if and how this process could work in YSO 
jets. Velocity fluctuations may possibly steepen into shocks and dissipate their energy
heating the gas, but that radiative losses come into play and 
act against the heating process.

An alternative explanation by L-FCD2000 to interpret the line ratios of DG Tau 
(see also Hartigan (2004)), is that one observes several post-shock regions of high excitation 
with a filling factor of $\sim$1\% (`shocking jet'). L-FCD2000 found that the line ratios 
of DG Tau and other HH objects (compilation by Raga et al. 1996) might be interpreted 
as series of shocks arrayed along the jet with varying shock velocities as one moves along the 
radius away from the jet axis, typically: 70-100 km s$^{-1}$ for the HV component, 
50-60 km s$^{-1}$ 
for IV and pre-shock densities that decrease from the star as $\propto r^{-2}$ starting from
$n_0=10^5$ cm$^{-3}$. In their analysis, they considered post-shock parameters consistent with
planar, stationary shocks models (e.g. Hartigan et al. 1994).  

Raga et al. (2001) modeled the morphological and dynamical properties of the DG Tau jet
by means of 3D numerical simulations, assuming a precessing jet with a velocity that varied
sinusoidally in time.
In the present paper we restrict ourselves to the interpretation of the line intensity ratio
behavior
along the jet, abandoning the assumption of stationary shocks. We follow the
dynamical evolution of an initial perturbation as it steepens into a (radiative) shock traveling
along the jet, and derive the post-shock physical parameters consistently (Massaglia et al. 2005a).   
From these parameters we construct the emission line ratios to be compared with observations.
In this first analysis we consider the evolution of a single shock, neglecting the possible 
interaction with other shocks.

The plan of the paper is the following: in Section 2 we outline the initial conditions
and the computational scheme adopted; in
Section 3 we examine the shock evolution, while
in Section 4 we discuss the results and make comparisons with observations. Our conclusions are drawn
in Section 5.

\section{The model}
%

\subsection{Basic equations}
%
%

We restrict our attention to one-dimensional
MHD planar flow. The fluid is described in terms of
its density $\rho$, velocity $u$, thermal pressure $p$ and
(transverse) magnetic field $B_y$.  
Conservation of mass, momentum, magnetic field 
and total energy readily follows:  

\begin{equation}\label{eq:cont}
 \pd{\rho}{t} + \pd{(\rho u)}{x} = 0  \; ,
\end{equation}
\begin{equation}\label{eq:mom}
 \pd{(\rho u)}{t} + \pd{}{x}\left( \rho u^2 + p + \frac{B^2_y}{2}\right) = 0 \; ,
\end{equation}
\begin{equation}\label{eq:ind}
 \pd{B_y}{t} + \pd{(B_yu)}{x} = 0 \; ,
\end{equation}
\begin{equation}\label{eq:ener}
\frac{\partial E} {\partial t} + 
  \pd{}{x}\left[\left(E + p + \frac{B^2_y}{2}\right)u\right] = 
  - {\cal L}(T, f_{\rm n})\;,
\end{equation}
where $E = p / (\Gamma -1) + \rho u^2/2 + B_y^2/2$ is the total 
energy density (we use $\Gamma = 5/3$) and ${\cal L}(T, f_{\rm n})$ represents 
the energy loss term (energy per unit volume per unit time) which depends
on the temperature $T$ and, as explained below, on the number fraction of 
neutral hydrogen atoms, $f_{\rm n}$. 
The loss term accounts for energy lost in lines and in the ionization
and recombination processes. Line emissions include contributions  
from nine elements whose abundances have been assumed to be solar
 (Ansers \& Grevesse 1989): 
$H$ and $He$ resonance lines, the 13 strongest forbidden
lines of $C$, $N$, $O$, $S$, $Si$, $Fe$ and $Mg$.
$He$ is neutral whereas metals are singly ionized and the abundance of $C$ 
is 10\% of the solar one. 
The ionization states of $N$ and $O$ are fixed to that of $H$ by charge transfer.

To compare between the observed and computed line 
ratios, we have computed the populations of the atomic levels for
the forbidden transitions [SII]$\lambda
\lambda 6716, 6731$, [NII]$\lambda 6583$ and [OI]$\lambda 6300$, 
solving, according to Osterbrock (1974), the excitation - de-excitation equilibrium
equations for five energy levels.

An additional evolutionary equation is solved for  
the neutral fraction $f_{\rm n}$, 
\begin{equation}\label{eq:fneut}
\frac{\partial f_{\rm n}}  {\partial t} + 
 v \pd{f_{\rm n}}{x} =
n_{\rm e} [-c_{\rm i} f_{\rm n} + c_{\rm r} (1- f_{\rm n})] 
\;,
\end{equation}
where $n_{\rm e}$ is the electron density, whereas $c_{\rm i}$ and
$c_{\rm r}$ are, respectively, the ionization and recombination rate 
coefficients (Rossi et al. 1997). In this framework we have
\begin{equation}
n_{\rm e} = n_{\rm H} (1 - f_{\rm n}) + Z n_{\rm H} \;,
\end{equation}
and
\begin{equation}
p = n_{\rm H} (2 - f_{\rm n} + Y+ 2Z) k_{\rm B} T \;,
\end{equation}
where $n_{\rm H}$ is the total hydrogen density, $Y$ and $Z$ ($=0.001$) are
the Helium and metal abundances by number respectively. 
 We will also define the ionization fraction as
\begin{equation}
f_{\rm i} = \frac{n_{\rm e}} {n_{\rm H}} \;.
\end{equation}

\subsection{The initial perturbation}
%
%

A nonuniform pre-shock density that decreases away from the star is a reasonable 
ingredient when dealing with an expanding jet, as suggested by observations. 
Moreover, shocks that propagate into a stratified medium
tend to increase their amplitude when they find a decreasing pre-shock density.
Therefore, we consider the following nonuniform pre-shock density:
\begin{equation}
\rho_0(x)=\rho_0 \frac{x_0^2}{x_0^2 + x^2} \,.
\end{equation}
where $x$ is the spatial coordinate along the jet axis. 
Thus, for $x \gg x_0$, we have a conical decrease of density, i.e. a conical 
expansion of the jet, while for $x \lapp x_0$ the density decreases parabolically,
$\rho_0(x) \approx \rho_0 (1-x^2/x_0^2)$. $x_0$ sets the initial steepness of the density
function and this affects the shock evolution even at larger distances.

\begin{figure}
\resizebox{\hsize}{!}{\includegraphics{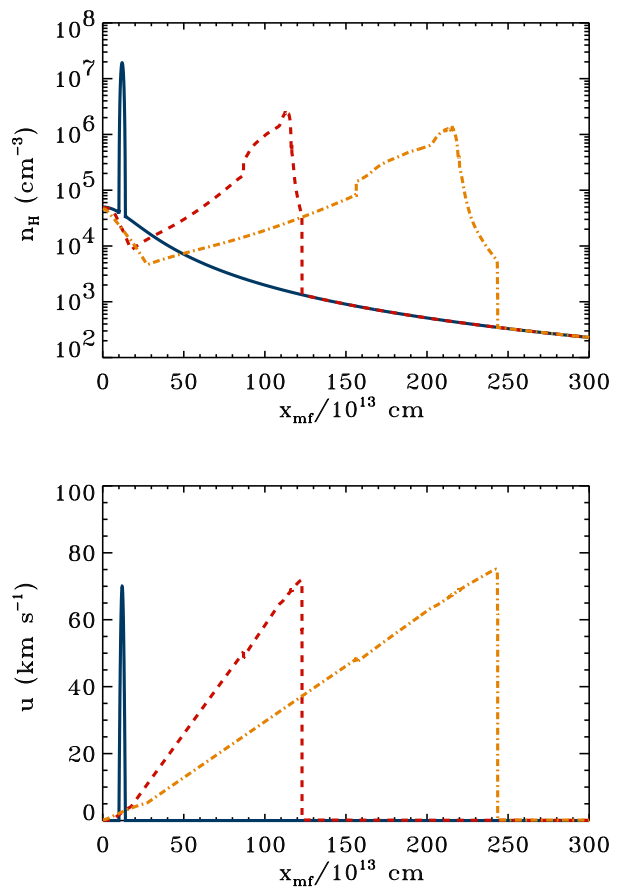}}
\caption{Density (upper panel) and velocity (lower panel) vs distance in the
reference frame at rest with the mean flow  (abscissa $x_{\rm mf}$); solid lines show the initial perturbation,
the dashed lines the evolution at $t=5$ ys and dot-dashed
lines at $t=10$ ys 
          }
\label{fig:pert}
\end{figure}

A generic initial perturbation imposed above the mean flow typically evolves forming two
shocks: the forward and reverse shocks (Hartigan \& Raymond 1993, Massaglia et al. 2005a).
The energy content of the perturbation therefore splits into two radiating shocks that 
propagate, decreasing their strengths along the way. Thus one cannot reach and maintain
the compression factors to explain observations,  as
happened in Massaglia et al. (2005a), unless assuming an initial perturbation
velocity amplitude of the order of the jet mean flow speed. 
Thus it would be more desirable to work with a single forward shock. 
To this purpose, we consider a disturbance that maintains the Riemann invariant 
$J_-$ constant  (Zeldovich \& Raizer 1966):
\begin{equation}
J_-=u-\int \frac{dp}{\rho c} = u-\int \frac{dp}{du} \frac{du}{\rho c} = {\rm constant} \;.
\label{eq:riem}
\end{equation}
After differentiation and some algebra, one finds
\begin{equation}
\rho=\left[ \frac{\gamma-1}{2 \sqrt{K\gamma}} (u-U_0)+\rho_0^{\frac{\gamma-1}{2}} \right]^{\frac{2}{\gamma-1}}  \;.
\label{eq:riem4}
\end{equation}
We set $U_0(x)=0$, implying that we carry out our calculations 
in the reference frame of the mean flow. We will have to
transform our data to the laboratory frame when comparing results with
observations. We prescribe the velocity perturbation as
\begin{displaymath}
u(x)=\left\{ \begin{array}{ll}
u_0 [-(x-x_{\rm 1})^2 + 2 \sigma (x-x_{\rm 1})] &  \textrm{if \ $2 \sigma + x_{\rm 1}> x>x_{\rm 1}$ } \\
0 & \quad \; \textrm{otherwise}
\end{array} \right.
\end{displaymath}
where $ x_{\rm 1}$ is the initial coordinate of the perturbation 
and $\sigma$ is its half-width (see Fig. \ref{fig:pert}, solid line).
Different choices of the initial perturbation shape are not
crucial for the shock formation and evolution.

 Notice that, strictly speaking, $J_-$ in Eq. \ref{eq:riem} is no longer an 
hydrodynamic invariant when magnetic fields are present. 
In our case, however, the initial magnetic pressure is typically 
small and the resulting perturbation still produces a strong 
forward shock, as shown in Fig. \ref{fig:pert}.
In our calculations we have assumed $x_1=10^{14}$ cm and $\sigma=2 \times 10^{13}$ cm.

The boundary conditions assume free outflow at $x=0$ and 
$x=L$.
The extent of the computational domain, $L$, has been chosen 
sufficiently large to follow the shock evolution for 
$t \sim 15$ ys and avoid spurious interactions with the boundaries. 
For this reason we adopt  $L = 4.5 \times 10^{15}$ cm.  

%

\subsection{Method of solution}
%
%

To solve the MHD equations we have employed the hydrodynamical code
PLUTO (Mignone, Massaglia \& Bodo, 2004). 

Equations (\ref{eq:cont})--(\ref{eq:ener}), together with 
Eq. (\ref{eq:fneut}) are solved using a conservative, second-order
accurate total variation diminishing (TVD) scheme. 
Piecewise linear interpolation with limited slopes is used 
to ensure monotonicity inside each computational zone.
Second order accuracy in time is achieved via a MUSCL-Hancock 
predictor step (van Leer, 1985). A linear, Roe-type, Riemann 
solver is used to compute the inter-cell fluxes needed 
in the conservative update.  
The conservative formulation is essential for a correct description
of the shock dynamics (LeVeque et al., 1998).
Source terms describing cooling, ionization and recombination 
processes are treated using operator splitting.

In the simulations presented below, the region of interest is confined 
mainly to the post-shock flow, where most of the emission takes place.
Since this region is much smaller in size ($\lapp \ 7 \times 10^{13}$ cm) than the
domain and the shock front is not stationary, a static uniform 
grid would demand increasingly high resolution in order to adequately
resolve the post-shock structure. 
For this reason, we use an Adaptive Mesh Refinement (AMR) technique (described in
Massaglia et al. 2005b) when solving Eqs. (\ref{eq:cont})--(\ref{eq:ener}),
thus reaching an effective resolution of about $360,000$ grid points.
 
This allows to considerably speed up the computation with 
sufficient resolution to accurately describe the dynamics and 
emission processes.

\section{Results and discussion}
%

\begin{figure}
\resizebox{!}{!}{\includegraphics[scale=0.4]{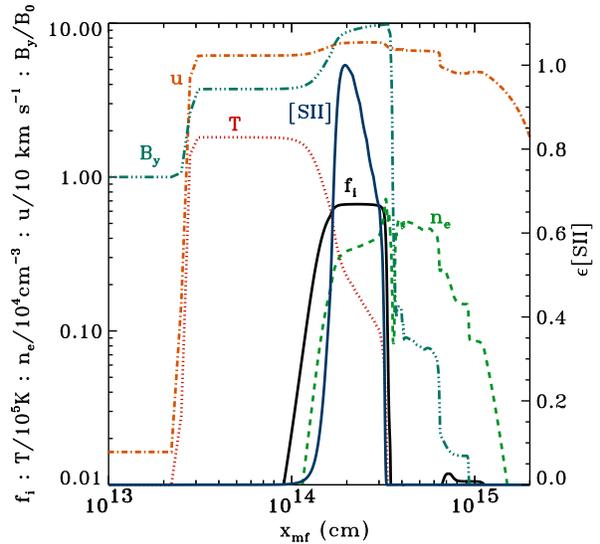}}
\caption{Spatial behavior, in the mean flow frame, of the physical quantities as they 
undergo the shock transition, from left to right, after 10ys of evolution, corresponding
to a traveled distance of the shock front of about $6.5 \times 10^{15}$ cm along the jet axis. Labels are:
 $f_{\rm i}$ ionization fraction (solid line), $n_{\rm e}$ electron density (dashed line), $T$ temperature (dotted line), $u$ velocity (dot-dashed line),
 $B_y$ magnetic field (in units of the pre-shock value $B_0$, dot-dot-dashed line) and $[SII]$ emissivity (solid line)
          }
\label{fig:post}
\end{figure}

We have followed the temporal evolution of the perturbation as it steepened 
into a shock traveling along the computational domain (see Fig. \ref{fig:pert}). 
We have been able, with the AMR technique employed, to resolve the post-shock
region at all times and in Fig. \ref{fig:post} we show  the post-shock variables 
versus the spatial coordinate.
Our model calculations depend on the following parameters: 
the initial perturbation velocity amplitude $u_0$, the pre-shock magnetic field
transverse to the flow $B_0$, the pre-shock density parameters $x_0$
and $n_0$ and the mean flow speed $U_0$, to transform the results 
back to the observer's reference frame. 
Test calculations have shown that, among these parameters, $u_0$ and 
the pre-shock density parameters $x_0$ and $n_0$ are the most critical 
for the final results. We have fixed the 
(isothermal) pre-shock temperature $T=1,000$ K (Raga et al. 2001) and 
ionization fraction  $f_{\rm i}=Z$ due to metals only.
However results are weakly sensitive to these values, for the range of parameters considered.
In Fig. \ref{fig:post} we plot the post-shock temperature $T$, electron density ($n_{\rm e}$), 
ionization fraction ($f_{\rm i}$),  magnetic field $B_y$ in units of $B_0$, 
[SII] emissivity (in units of the maximum value, rightmost vertical scale)
and velocity  $u$, in the reference frame of the mean flow. The horizontal
scale has an arbitrary origin and tells us the size of the post-shock region. 
The values of the parameters are: $u_0=70$ km s$^{-1}$,
$B_0=100 \ \mu$G, $x_0=0^{\prime \prime}.1$ and $n_0=5 \times 10^4$ cm$^{-3}$. 

The post-shock quantities are shown after 10 ys of evolution, 
when the shock front has traveled for about 
 $6.5 \times 10^{15}$ cm from the source, having assumed a mean flow speed $U_0=150$ km s$^{-1}$ (see below).
We note that the post-shock temperature attains a value $\gapp \
10^5$ K right behind the shock and decreases to below $10^4$ K about $7 \times 10^{13}$ cm 
from the shock front, where the [SII] emission takes place  and the magnetic field
reaches its maximum and subsequently drops to low values at $\sim 10^{14}$ cm. 
We see that both the (collisional) ionization fraction and the electron density 
reach maximum values at about $7 \times 10^{13}$ cm behind the shock front. 
In the post-shock region the particle velocity remains nearly
constant to the original value of the perturbation amplitude 
and decreases  starting at $\sim 2 \times 10^{15}$ cm from the shock front. 
We are {\it not} plotting the post-shock speed in the frame where the shock is at rest.
 At earlier times, the general behavior
would be almost unchanged for each quantity exception made for the electron density 
which would present a maximum that is higher by about a factor of three due to the higher 
(imposed) pre-shock density. Clearly, the opposite happens at later times in 
the evolution.

In order to carry out a comparison with observations we need to average the post-shock quantities
at every evolutionary time point. We have done this following
Hartigan et al. (1994), i.e. defining the $[SII]$-weighted average as follows
\[
\langle Q \rangle= \frac{\int Q(x) \epsilon \{[SII](x)\} \ dx}
{ \int \epsilon \{[SII](x)\} \ dx}
\]
where $Q$ is a physical quantity such as  either electron density or ionization fraction or
the line emissivity ratio of the sulfur doublet $[SII]\lambda 6716/[SII]\lambda 6731$.
Concerning the line emissivity ratio $[NII]\lambda 6583/[OI]\lambda 6300$, we have carried out the averaging
procedure in the same way, but adopting the total emissivity $\epsilon \{[NII](x)\}+\epsilon \{[OI](x)\} $
as the weighting function. However, we noticed that the particular choice of the weighting
function has very little effect on the results.  Note also from Fig. \ref{fig:post} that
for $f_{\rm i}$ and $B_y$ the averaging procedure will yield, with a good
approximation, the maximum value of these quantities.

\begin{figure}
\vskip 20pt
\resizebox{\hsize}{!}{\includegraphics{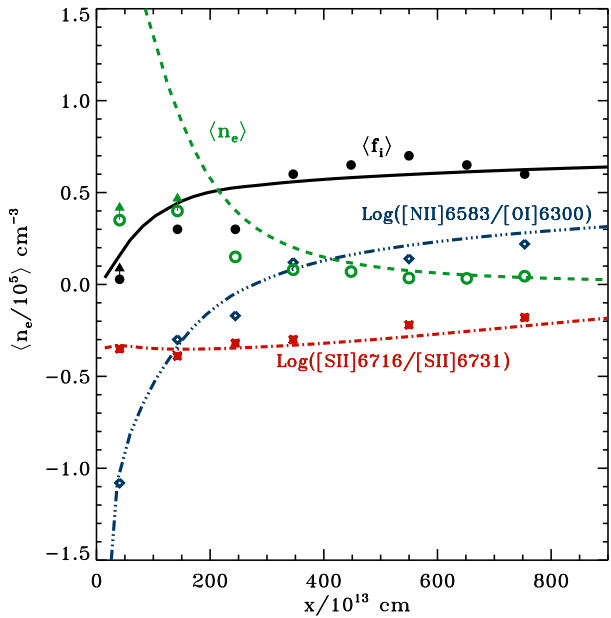}}
\caption{Averaged values of the post-shock electron density ($\langle n_{\rm e} \rangle$, dashed
line), ionization fraction ($\langle f_{\rm i} \rangle$ solid line), ratios of [SII] (dot-dashed line)
and [NII]/[OI] (dot-dot-dashed line) along the jet for the reference parameters vs distance
(in the laboratory frame): 
$u_0=70$ km s$^{-1}$, $B_0=100 \ \mu$G, $x_0=
0^{\prime \prime}.1$ and $n_0=5 \times 10^4$ cm$^{-3}$ (see also Fig. \ref{fig:post}).
 Here we have set $U_0=150$ km $s^{-1}$
          }
\label{fig:evol}
\end{figure}

In Fig. \ref{fig:evol} we compare of the averaged post-shock quantities, obtained
for the above set of parameters and assuming a mean outflow velocity $U_0=150$ km s$^{-1}$,
with the observations by L-FCD2000. The abscissa indicates the position
of the post-shock region in the observer's frame, $x=U_0 t + \xi(t)$, where $t$ is the elapsed time 
and $\xi(t)$ is the location of the emission region. We have chosen the high
velocity (HV) jet data of the line ratios, ionization fraction and
electron density from the paper of L-FCD2000, likely corresponding to the inner
part of the jet about its longitudinal axis. After examining Fig. \ref{fig:post}, we have estimated $\xi(t)$ 
as the shock front position minus an interval $d$
($=2 \times 10^{14}$ cm) that accounts for the shift of the emission region behind the
shock front. The results are nearly insensitive to the choice of $d$.
We recall that the {\it directly} observable data are 
the line intensity ratios (in Fig. \ref{fig:evol}, crosses are
 for the [SII] doublet and diamonds for
the [NII]/[OI] ratio), while the ionization fraction and electron density have been obtained 
by L-FCD2000 adopting the diagnostic technique described in Bacciotti et al (1995). 
From Fig. \ref{fig:evol} we note that the [NII]/[OI] and [SII] ratios fit the data
remarkably well, 
while the electron density and the ionization fraction exceed the ones {\it derived} 
from observations at low values of the spatial coordinate $x$. 
 In fact, Bacciotti et al. (1995) obtain $n_{\rm e}$ from the sulfur doublet line 
intensity ratio, which saturates for electron densities higher than the critical density 
$n_{\rm c} \lapp \ 10^4$ cm$^{-3}$. This leads to an underestimation of the electron density and, 
consequently, the ionization fraction. Note that in L-FCD2000 the first couple of data points of 
the electron density and the first one of the ionization fraction are lower limits.
Instead, this method is appropriate for lower values of the electron density, thus the
agreement in Fig. \ref{fig:evol} at larger distances.

Several questions arise: which sets of parameters better represent the observations? 
What are the crucial parameters?

Variations of the advection velocity $U_0$ have no physical impact on the shock evolution but 
concern only a dilation of $x$ axis. 
 We varied $U_0$ by $\pm$ 40\% and noted that the agreement
remained reasonably good. Thus values of $U_0$ within these two limits 
still interpret observations reasonably well.

We are left with four main parameters: $n_0$, $x_0$, $B_0$ and $u_0$. 
Due to the  difficulty in carrying out a full exploration of the parameter space, 
we have synthesized the `goodness' of the agreement between
model curve and observational points of the line ratios defining a 
variance $\sigma^2=\sum_i{(O_i-E_i)^2}$, where $O_i$ are the data and $E_i$ the model points. 
In Table 1 we show the qualitative behavior of the [SII] and [NII]/[OI] ratio model curves 
when a parameter on the left column is increased, leaving the other ones unchanged with their 
reference values of Fig. \ref{fig:evol}: these curves are displaced downward when $n_0$ and $x_0$ are raised, changing in shape as well, while the curves shift upward when raising $B_0$. 
 We have run several models (according to Table 1) varying the parameters 
$n_0$, $B_0$, $u_0$ and $x_0$ around the reference set and calculated the variance 
$\sigma^2$ of the corresponding fits. 
Since for most of the spatial coordinate $x$ we are above the
critical density, the variance of the fits of the [SII] doublet line ratio is not very sensitive
to the changes in these parameters. However, this is not the case for the [NII]/[OI] curves:
we could obtain fits for [NII]/[OI] with a variance close to the minimum by varying 
$n_0$, $x_0$ and $B_0$  by about 30\% and $u_0$ by about 10\% plus or minus with respect
to the reference values.

As far as the magnetic field evolution is concerned, direct comparisons with previous 2D models
(e.g., Garcia et al. 2001a) and 3D simulations (e.g., Cerqueira \& de Gouveia dal Pino 2001) are 
difficult because we only include the transverse component of the magnetic field. 
The spatial evolution of the mean field in
our models typically shows a monotonic decrease from about $30 B_0$ down to about $5 B_0$. 
If we consider the behavior of $B_\phi$ from Garcia et al. (2001a), our values of the magnetic field,
with $B_0=100 \mu{\rm G}$, remain a factor of $\sim 2-3$ below the curve of their model C.

\begin{table} 
\begin{center}
\caption{Behavior of the line ratio curves with the increase of the parameter in the leftmost column} 
 \begin{tabular}{|c||c|c|} \hline  
 Parameter & [NII]6583/[OI]6300 &  [SII]6716/[SII]6731  \\ 
\hline \hline 
$n_0$ & $\downarrow$ &  $\downarrow$   \\ \hline 
$u_0$ & $\uparrow$  & $\downarrow$     \\ \hline
$x_0$ & $\downarrow$ &  $\downarrow$   \\ \hline 
$B_0$ & $\uparrow$   & $\uparrow$   \\ \hline 
\end{tabular}
\end{center}
 
\end{table} 

\section{Conclusions}
With the goal to interpret the line-emission within the first $5^{\prime\prime}$ observed in some
Herbig-Haro jets, 
we have considered the evolution of a strong planar perturbation in a stratified, magnetized medium and in the
presence of radiative losses. We have examined the temporal evolution of this perturbation as
it steepened into a shock and focused our attention to the structure of the post-shock region. 
Having set physical parameters consistent with the environment of stellar jets in their 
inner portions, closer to the young star, we have derived averaged line intensity ratios 
that could be compared with observations. We have adopted observations
of line ratios of the DG Tau jet by L-FCD2000, as an example, and found that even in this extremely 
simple model, one can interpret observations with a constrained set of parameters reasonably well. 

In this model we have  {\it a priori} prescribed the longitudinal density profile upstream of the perturbation,
assuming that a real jet, in the same physical conditions, would expand decreasing its density along the way. 
Whether an actual jet behaves in a similar fashion remains to be seen and, to this purpose,
2D MHD simulations of radiative jets are needed.
 The reference set of parameters that yield a `good'
agreement with observations are: $u_0=70$ km s$^{-1}$, 
$B_0=100 \ \mu$G, $x_0=0^{\prime \prime}.1$, $n_0=5 \times 10^4$ cm$^{-3}$ with a mean flow
speed $U_0=150$ km s$^{-1}$. Assuming an initial radius of $3 \times 10^{14}$ cm (Raga et al.
2001), these values are consistent with a mass loss rate $\dot M \approx 5 \times 10^{-8}$
M$_\odot$ ys$^{-1}$, in agreement with the estimates of L-FCD2000. The presence of a perturbation 
raises this value to $\dot M \approx 10^{-6}$ M$_\odot$ ys$^{-1}$ for short periods. 

One might ask whether these shocks may also be responsible for the emission knots observed in
some HH jets (e.g. HH34, HH111) at larger distances from the source ($\lapp \ 45^{\prime\prime}$)
(c.f. a review by Raga, Beck \& Riera 2004, Micono et al. 1998a,b). Due
to the strong emissivity of the shock compressed medium, we believe that these shocks would 
hardly be
visible at very large distances, as observed.

Thus, these calculations successfully reproduce line-ratio observations and thus
strongly support the hypothesis of L-FCD2000 of the DG Tau jet as a `shocking jet',
i.e. that one actually observes not a continuous emitting jet but just the gas parcels that have
undergone compression, heating and ionization in shocks.

\begin{acknowledgements}
We thank Sylvie Cabrit for stimulating suggestions and helpful comments. 
 We are also grateful to Pat Hartigan for discussions  and to Paulo Garcia who, as referee,
helped us to improve the paper.
We acknowledge the Italian MIUR for financial support, grants No. 2002.028843 and No. 2004.025227.
The present work was supported in part by the European Community Marie Curie
Actions - Human resource and mobility within the JETSET network under
contract MRTN-CT-2004 005592. 

\end{acknowledgements} 

{}

\end{document}